\documentclass[twocolumn,preprintnumbers,amsmath,amssymb]{revtex4}
 \usepackage{graphicx}

\begin{document}
 \preprint{}

\newcommand{\api}{$a^3\Pi$}
\newcommand{\X}{$X^1\Sigma^+$}

\title{Ramsey-type microwave spectroscopy on CO ($a^3\Pi$)}

\author{A. J. de Nijs}
\author{W. Ubachs}
\author{H. L. Bethlem}
\email{H.L.Bethlem@vu.nl}
\affiliation{Department of Physics and Astronomy, LaserLaB, VU University, De Boelelaan 1081, 1081 HV Amsterdam, The Netherlands}

\begin{abstract}
Using a Ramsey-type setup, the lambda-doublet transition in the $J=1,\, \Omega=1$ level of the \api\ state of CO was measured to be 394~064~870(10)~Hz. In our molecular beam apparatus, a beam of metastable CO is prepared in a single quantum level by expanding CO into vacuum and exciting the molecules using a narrow-band UV laser system. After passing two microwave zones that are separated by 50~cm, the molecules are state-selectively deflected and detected 1~meter downstream on a position sensitive detector. In order to keep the molecules in a single $m_J^B$ level, a magnetic bias field is applied. We find the field-free transition frequency by taking the average of the $m_J^B = +1 \rightarrow m_J^B = +1$ and $m_J^B = -1 \rightarrow m_J^B = -1$ transitions, which have an almost equal but opposite Zeeman shift. The accuracy of this proof-of-principle experiment is a factor of 100 more accurate than the previous best value obtained for this transition. 
\end{abstract}

\maketitle

\section{Introduction}

Measurements of transition frequencies in atoms and atomic ions nowadays reach fractional accuracies below 10$^{-17}$, making atomic spectroscopy eminently suitable for testing fundamental theories~\cite{Rosenband:science2008,Huntemann:prl2012,Nicholson:prl2012}. The accuracy obtained in the spectroscopic studies of molecules typically lags by more than three orders of magnitude, however, the structure and symmetry of molecules gives advantages that make up for the lower accuracy in specific cases. Molecules are for instance being used in experiments that search for time-symmetry violating interactions that lead to a permanent electric dipole moment (EDM) of the electron~\cite{Hudson:Nature2011,ACME2014}, tests on parity violation~\cite{Daussy:prl1999,Quack:fardisc2011}, tests of quantum electrodynamics~\cite{Dickenson:prl2013}, setting bounds on fifth forces~\cite{Salumbides:prd2013} and testing a possible time-variation of the proton-to-electron mass ratio~\cite{shelkovnikov:prl,Jansen:spotlight,bagdonaite:science2013,Truppe:natcomm2013}.

Here we present the result of high-resolution microwave spectroscopy on CO molecules in the metastable \api\ state. Metastable CO has a number of features that make it uniquely suitable for precision measurements; (i) CO is a stable gas that has a high vapor pressure at room temperature making it straight forward to produce a cold, intense molecular beam; (ii) The \api\ state has a long lifetime of $\sim$2.6~ms~\cite{Gilijamse} and can be excited directly to single rotational levels at a well-defined time and position using laser radiation around 206~nm~\cite{Drabbels:cpl1992}; (iii) The metastable CO can be detected position dependently on a microchannel plate detector~\cite{Jongma:JCP}, allowing for a simple determination of the forward velocity as well as the spatial distribution of the beam; (iv) The most abundant isotopologue of CO ($^{12}$C$^{16}$O, 99\% natural abundance) has no hyperfine structure, while other isotopologues are available commercially in highly enriched form; (v) The \api\ state has a strong first order Stark and Zeeman shift, making it possible to manipulate the beam using electric or magnetic fields~\cite{Gilijamse,Jongma:CPL,Bethlem:prl1999}.  

Recently, it was noted that the two-photon transition between the $J = 8,\, \Omega=0$ and the $J=6,\, \Omega=1$ levels in the \api\ state is 300 times more sensitive to a possible variation of the proton to electron mass ratio ($\mu$) than purely rotational transitions~\cite{Bethlem:FarDisc,denijs:pra2011}. We plan to measure these transitions with high precision. Here, as a stepping stone, we present measurements of the lambda-doublet transition in the $J=1,\, \Omega=1$ level around 394~MHz using Ramsey's method of separated oscillatory fields. 

\section{Energy level diagram}
\label{sec:mwlevelscheme}

\begin{figure}
\begin{center}
\includegraphics[width=\linewidth]{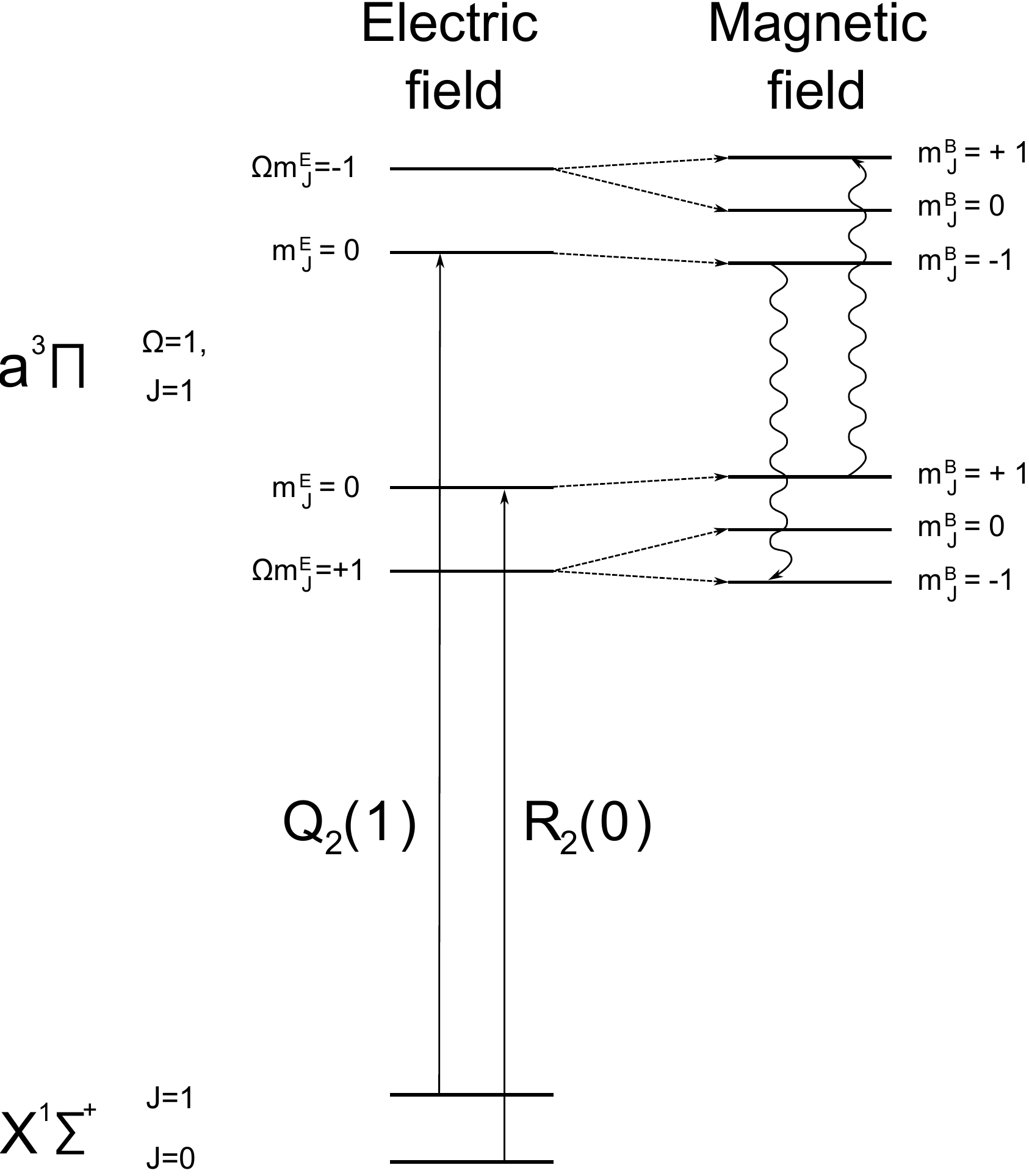}
\end{center}
\caption{Energy level diagram showing the levels relevant for this experiment. CO molecules are excited from either of the two lowest rotational levels of the \X\ ground state to the upper or lower lambda doublet component of the $\Omega=1, J=1$ level of the \api\ state using a narrow-band UV laser, indicated by the vertical arrows. An electric field is applied to lift the degeneracy of the $m_{J}^{E}$ sublevels, enabling the excitation of a single $m_{J}^{E}$ sublevel. In a magnetic field that is perpendicular to the electric field, the four $m_{J}^{E}$ sublevels correspond to six $m_{J}^{B}$ sublevels as indicated. The wavy arrows indicate the two microwave transitions that are measured in this work.}
\label{fig:lvlsc_mw}
\end{figure}

The \api\ state of CO is one of the most extensively studied triplet states of any molecule. The transitions connecting the \api\ state to the \X\ ground state were first observed by Cameron in 1926~\cite{Cameron}. Later, the \api\ state was studied using radio frequency~\cite{Freund,Wicke:1972,Gammon:1971}, microwave~\cite{Saykally:1987,Carballo,Wada}, infrared~\cite{Havenith:MolPhys1988,Davies}, optical~\cite{Effantin:1982} and UV spectroscopy~\cite{Field}. We have recently measured selected transitions in the CO \api\ - \X\ (0-0) band using a narrow-band UV laser~\cite{denijs:pra2011} resulting in a set of molecular constants that describes the level structure of the \api\ state with an absolute accuracy of 5~MHz with respect
to the ground state and a relative accuracy of 500~kHz within the \api\ state.

Fig.~\ref{fig:lvlsc_mw} shows the levels relevant for this study. The CO molecules are excited to the $\Omega=1$ manifold of the \api\ state from the $J=0$ and $J=1$ levels of the \X\ state using either the R$_2$(0) or the Q$_2$(1) transitions, indicated by straight arrows. The excitation takes place in an electric field that splits both lambda-doublet components into two levels labeled by $\Omega m_J^E$ as shown on the left hand side of the figure. The microwave transition is recorded in a region that is shielded from electric fields, but that is subjected to a homogeneous magnetic field. In a magnetic field, both lambda-doublet components are split into three levels labeled by $m_J^B$ as shown on the right hand side of the figure. In the region between the excitation zone and the microwave zone the applied magnetic field is perpendicular to the electric field. In this case, the $m_J^E=0$ sublevel of the upper lambda-doublet component corresponds to the $m_J^B=-1$ sublevel while the $m_J^E=0$ sublevel of the lower lambda-doublet component corresponds to the $m_J^B=+1$ sublevel, as indicated by the dashed arrows~\cite{Meek:pra2011}. The $\Omega m_J^E=+1$ and $\Omega m_J^E=-1$ sublevels correspond to the $m_J^B=0,-1$ and $m_J^B=0,+1$ sublevels, respectively. The $m_J^B=+1$ and $m_J^B=-1$ sublevels exhibit a linear Zeeman effect of $\sim$1~MHz/Gauss, respectively, while the $m_J^B=0$ sublevel does not exhibit a linear Zeeman effect. Ideally, we would therefore record the $m_J^B=0 \rightarrow m_J^B=0$ transition. However, this transition is not allowed via a one-photon electric or magnetic dipole transition. Instead, we have recorded the $m_J^B=+1 \rightarrow m_J^B=+1$ and $m_J^B=-1 \rightarrow m_J^B=-1$ transitions indicated by the wavy arrows in the figure. To a first-order approximation, these transitions do not display a linear Zeeman shift. However, the mixing of the different $\Omega$ manifolds is parity dependent and as a result the Zeeman shift in the upper and lower lambda-doublet components are slightly different. Hence, the $m_J^B=+1 \rightarrow m_J^B=+1$ and $m_J^B=-1 \rightarrow m_J^B=-1$ transitions show a differential linear Zeeman effect of $\sim$10~kHz/Gauss. This differential linear Zeeman effect is opposite in sign for the two recorded transitions, and is canceled by taking the average of the two.  

\section{Experimental Setup}
\label{sec:mwexperimentalsetup}

\begin{figure*}
\begin{center}
\includegraphics[width=\linewidth]{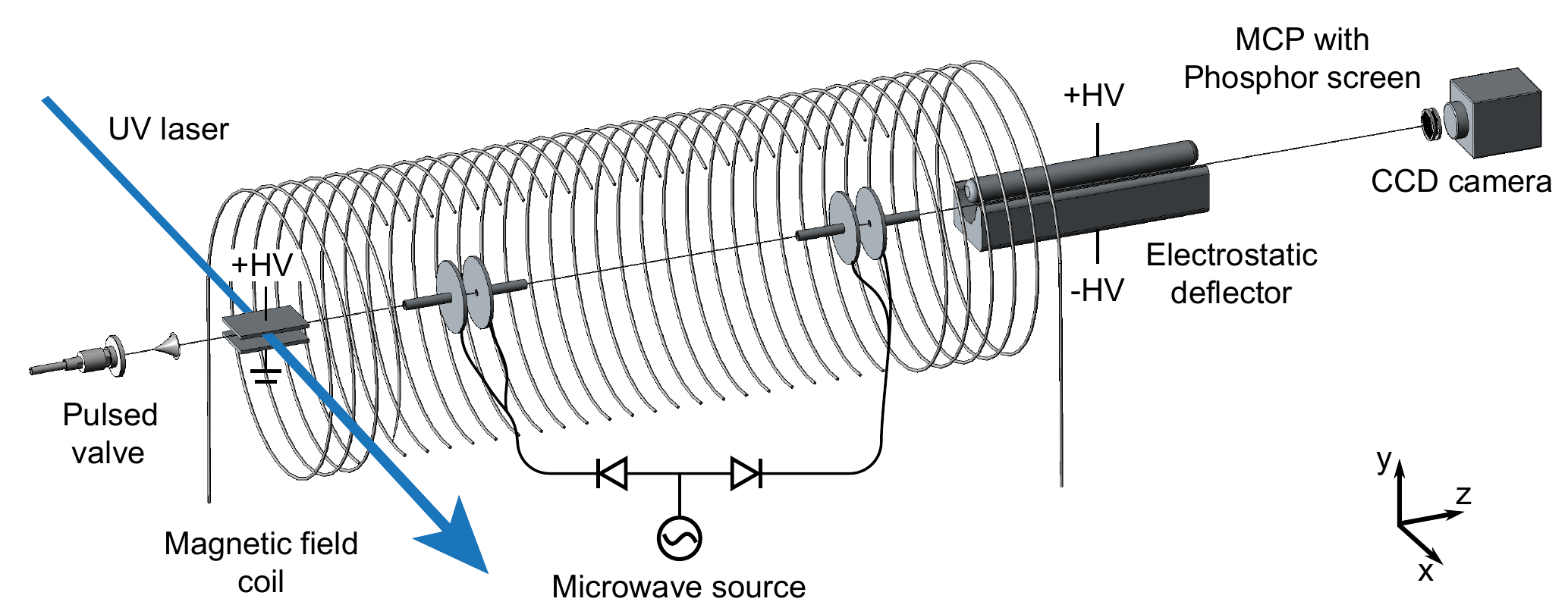}
\end{center}
\caption{Schematic drawing of the molecular beam setup. A supersonic, rotationally cold beam of CO molecules is produced by expanding CO gas into vacuum using a solenoid valve. After passing a 1~mm diameter skimmer, the molecules are excited to the metastable \api\ state using laser radiation tunable near 206~nm. An electric field of 1.5~kV/cm is applied in the excitation zone. Perpendicular to this electric field, a bias magnetic field of typically 17~Gauss is applied. Microwave transitions are induced by two microwave zones separated by 50~cm, in a Ramsey type setup. After passing the second microwave zone, molecules are state-selectively deflected using an inhomogeneous electric field and are subsequently detected using a position sensitive detector.}
\label{fig:mw_setup}
\end{figure*}

Fig.~\ref{fig:mw_setup} shows a schematic drawing of the molecular beam-machine used in this experiment. A supersonic, rotationally cold beam of CO molecules is produced by expanding either pure CO or a mixture of 20\% CO in He into vacuum, using a solenoid valve (General Valve series 9). The backing pressure is typically 2 bar, while the pressure in the first chamber is kept below 10$^{-5}$ mbar. After passing a 1~mm skimmer, the molecular beam is  crossed at right angles with a UV laser beam that excites the molecules from the \X\ ground state to the \api\ state. Details of the laser system are described elsewhere~\cite{Hannemann:PRA1,Hannemann:rsi2007,denijs:pra2011}. Briefly, a Ti:sapphire oscillator is locked to a CW Ti:sapphire ring laser and pumped at 10~Hz with a frequency doubled  Nd:YAG-laser. The output pulses from the oscillator are amplified in a bow-tie type amplifier and consecutively doubled twice using BBO crystals. Ultimately, 50~ns, 1~mJ pulses around 206~nm are produced, with a bandwidth of approximately 30~MHz. 

In the laser excitation zone a homogeneous electric field of 1.5~kV/cm is applied along the $y$-axis which results in a splitting of $\sim$500~MHz between the $\Omega m_J^E=\pm1$ and $m_J^E=0$ sublevels, large compared to the bandwidth of the laser. In addition, a homogeneous magnetic field of typically 17~Gauss is applied along the molecular beam axis by running a 2~A current through a 100~cm long solenoid consisting of 600 windings. The current is generated by a current source (Delta electronics ES 030-10) that is specified to a relative accuracy of 10$^{-3}$. In the absence of an electric field, the magnetic field would give rise to a splitting of the $m_J^B$ sublevels of $\sim$15~MHz. In a strong electric field perpendicular to the magnetic field, the orientation and $m$ labeling is determined by the electric field and the magnetic field splitting is below 1~MHz, small compared to the bandwidth of the laser. In our experiments, we excite the molecules to the $m_J^E=0$ sublevel of either the lower or upper lambda-doublet component via the R$_2$(0) or the Q$_2$(1) transitions using light that is polarized in the $z$ or $y$-direction, respectively (see for axis orientation the inset in Fig.~\ref{fig:mw_setup}). Upon exiting the electric field, the $m_J^E=0$ sublevel of the lower lambda-doublet component adiabatically evolves into the $m_J^B=+1$ sublevel, while the $m_J^E=0$ sublevel of the upper lambda-doublet component adiabatically evolves into the $m_J^B=-1$ sublevel, see Fig.\ref{fig:lvlsc_mw}.

The microwave measurements are performed in a Ramsey-type setup consisting of two microwave zones that are separated by $\sim$50~cm. Each microwave zone consists of two parallel cylindrical plates spaced 20~mm apart, oriented perpendicularly to the molecular beam axis with 5~mm holes to allow the molecules to pass through. Tubes of 5~cm length with an inner diameter of 5~mm are attached to the field plates. The two interaction regions are connected to a microwave source (Agilent E8257D) that generates two bursts of 50~$\mu$s duration such that molecules with a velocity close to the average of the beam are inside the tubes when the microwave field is turned on and off. The electric component of the microwave field is parallel to the bias magnetic field along the $z$-axis, allowing $\Delta m_J^B=0$ transitions only. Directional couplers are used to prevent reflections from one microwave zone to enter the other. Grids, not shown in the figure, are placed upstream and downstream from the interrogation zone to shield it from external electric fields.

After passing the microwave zones, the molecules enter a 30~cm long electrostatic deflection field formed by two electrodes separated by 3.4~mm, to which a voltage difference of 12 or 20~kV is applied for pure CO and CO seeded in helium, respectively. Ideally, the deflection field exerts a force on the molecules that is strong and position-independent in the $y$-direction while the force in the $x$-direction is zero. It has been shown by de Nijs and Bethlem~\cite{denijs:pccp2011} that such a field is best approximated by a field that contains only a dipole and quadrupole term while all higher multipole terms are small. The electric field is mainly directed along the $y$-axis, i.e. parallel to the electric field in the laser excitation region. Hence, molecules that were initially prepared in a $m_J^E=0$ sublevel and have not made a microwave transition will not be deflected. In contrast, molecules that made a microwave transition will end up either in the $\Omega m_J^E=+1$ or $\Omega m_J^E=-1$ sublevels and will be deflected upwards or downwards, respectively.  

Finally, after a flightpath of 80~cm, sufficient to produce a clear separation between the deflected and non-deflected molecules, the molecules impinge on a Microchannel Plate (MCP) in chevron configuration, mounted in front of a fast response phosphor screen (Photonis P-47 MgO). The phosphor screen is imaged using a Charged-Coupled Device (CCD) camera (PCO 1300). Molecules in the \api\ state have 6~eV of internal energy which is sufficient to liberate electrons, that are subsequently multiplied to generate a detectable signal. The quantum efficiency is estimated to be on the order of 10$^{-3}$~\cite{Jongma:JCP}. The voltage on the front plate of the MCP is gated, such that molecules with a velocity of $\pm$3\% around the selected velocity are detected only, and background signal due to stray ions and electrons is strongly suppressed. The recorded image is sent to a computer that determines the total intensity in a selected area. A photomultiplier tube is used to monitor the integrated light intensity emitted by the phosphor screen. This signal is used for (manually) correcting the frequency of the UV laser if it drifts away from resonance. Note that due to the Doppler shift in the UV transition, the beam will be displaced in the $x$-direction when the laser frequency is off resonance. As the flight path of the molecules through the microwave zones will be different for different Doppler classes, this may result in a frequency shift. In future experiments this effect may be studied by measuring the transition frequency while changing the area of the detector over which the signal is integrated. At the present accuracy this effect is negligible.

\section{Experimental results}
\label{sec:mwresults}

\begin{figure}
\begin{center}
\includegraphics[width=\linewidth]{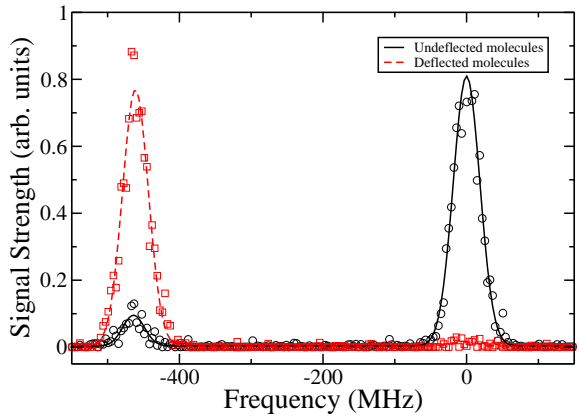}
\end{center}
\caption{Frequency scan of the UV-laser around the R$_2$(0) transition showing two peaks corresponding to the $\Omega m_J^E=1$ (left-hand side) and $m_J^E=0$ (right-hand side) sublevels, respectively. The black circles are recorded by integrating the signal of the undeflected beam, while the red squares are recorded by integrating the signal of the upwards deflected beam.}
\label{fig:deflecscan}
\end{figure}

Fig.~\ref{fig:deflecscan} shows the integrated intensity on the MCP detector as a function of the frequency of the UV laser tuned around the R$_2$(0) transition. An electric field of 1.5~kV/cm is applied along the $y$-axis while a magnetic field of 17~Gauss is applied along the $z$-axis. The polarization of the laser is parallel to the $y$-axis. The black curve shows the number of undeflected molecules while the red dashed curve shows the number of upwards deflected molecules. As the Stark effect in the \X\ state is negligible, the frequency difference between the two observed transitions directly reflects the splitting between the $\Omega m_J^E=+1$ and $m_J^E=0$  (lower lambda-doublet) sublevels in the $\Omega=1,J=1$ level of the \api\ state. The two peaks are separated by $\sim$500~MHz, as expected in the applied fields~\cite{Jongma:CPL}. When a magnetic field of 17~Gauss is applied, about 98\% of the molecules that were initially prepared in the $m_J^E=0$ sublevel remain in this sublevel until entering the deflection field, while 2\% of the molecules are non-adiabatically transferred to one of the $\Omega m_J^E=+1$ sublevels. At lower magnetic fields, the depolarization increases strongly. When no magnetic field is applied, only one third of the beam remains in the $m_J^E=0$ sublevel. Note, that although the depolarization gives rise to a loss in signal, it does not give rise to background signal in the microwave measurements.

\begin{figure*}
\begin{center}
\includegraphics[width=\linewidth]{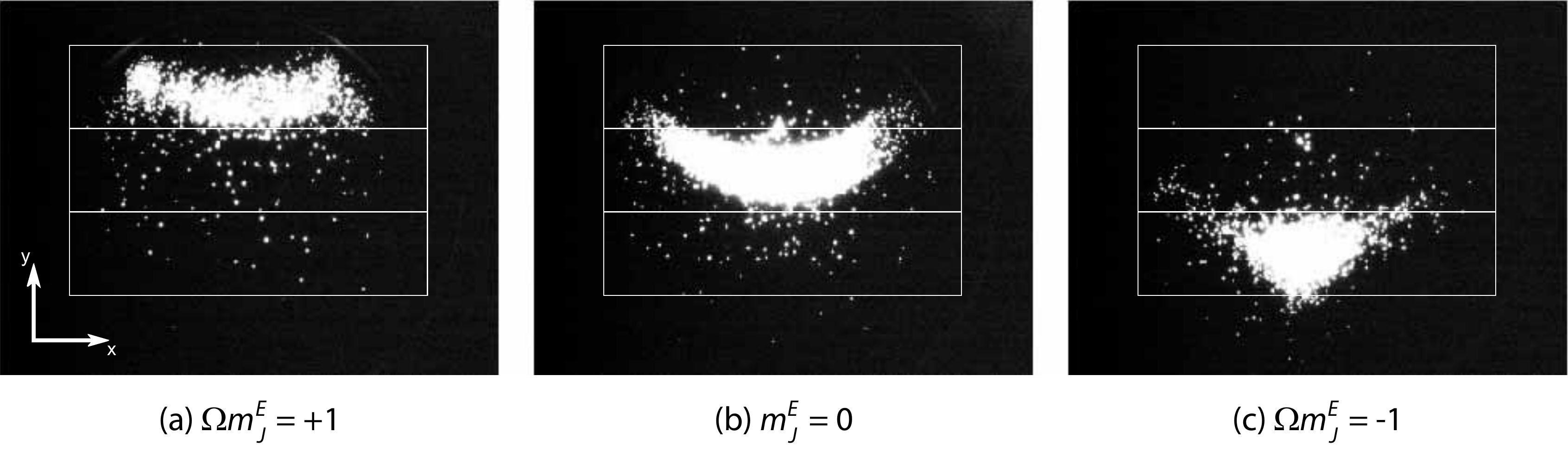}
\end{center}
\caption{Typical images recorded on the CCD camera showing the spatial distribution of the beam when the laser is resonant with transitions to different sublevels, as indicated. Molecules in the $\Omega m_J^E=+1$ and $\Omega m_J^E=-1$ sublevels are upwards or downwards deflected, respectively, while molecules in the $m_J^E=0$ sublevel experience almost no force. The white boxes define the integration areas used in the analysis.}
\label{fig:ima1}
\end{figure*}

Fig.~\ref{fig:ima1} shows a number of typical images recorded on the CCD camera when the frequency of the laser is resonant with a transition to (a) the $\Omega m_J^E=+1$ sublevel, (b) the $m_J^E=+0$ sublevel and (c) the $\Omega m_J^E=-1$ sublevel of the \api\ state. In this measurement, the exposure time of the CCD camera is set to be 1~s, i.e., each image is the sum of 10 shots of the CO beam. Each white spot in the image corresponds to the detection of a single molecule. As seen, molecules in the $\Omega m_J^E=+1$ sublevel are being deflected upwards, while at the same time they experience a slight defocusing effect in the $x$ direction. In contrast, molecules in the $\Omega m_J^E=-1$ sublevel are being deflected downwards, while experiencing a slight focusing effect in the $x$ direction. These observations are in agreement with the analysis given in de~Nijs and Bethlem~\cite{denijs:pccp2011}. The white boxes also shown in the figures define the area over which is integrated to determine the upwards deflected, undeflected and downwards deflected signal. Note that for recording these images the Ramsey tube containing the two microwave zones have been taken out. In this situation, typically, 1000~molecules are detected per shot. When the microwave zones are installed, the number of molecules reaching the detector is decreased by about a factor of 5. Gating the detector to select a specific velocity reduces the number of detected molecules further by a factor of 4. 

\begin{figure}
\begin{center}
\includegraphics[width=\linewidth]{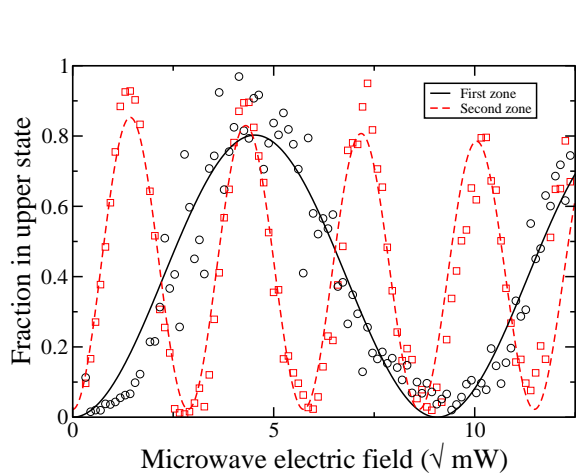}
\end{center}
\caption{Power dependence of the $m_J^B=+1 \rightarrow m_J^B=+1$ lambda-doublet microwave transition at resonance. The signal corresponds to the ratio of the number of molecules in the initial and final states. The black circles and red squares are recorded using the first and second microwave zone, respectively, while the solid and dashed curves result from a fit to the data.}
\label{fig:rabi}
\end{figure}

Fig.~\ref{fig:rabi} shows a power dependence of the microwave transition from the $m_J^B=+1$ sublevel in the lower lambda-doublet component to the $m_J^B=+1$ sublevel in the upper lambda-doublet component of the $\Omega=1,J=1$ level in the \api\ state. The frequency of the microwave field is set to the peak of the resonance. The signal corresponds to the ratio of the integrated signal over the boxes for the undeflected and deflected beam shown in Fig.~\ref{fig:ima1}. Note that in this way, any pulse-to-pulse fluctuations in the signal due to the valve and UV-laser are canceled. Typically, 50 molecules per shot are detected. The black circles and red squares are recorded using the first and second microwave zone, respectively. The curves also shown are a fit to the data using 

\begin{equation}
F(P)=a_0 \exp(-a_1\sqrt{P}) \sin^2(a_2 \sqrt{P}),
\end{equation}

\noindent
with $P$ being the microwave power. The observed deviations between the data and the fit are attributed to the fact that a fraction of the molecules that are deflected hit the lower electrode and are lost from the beam. As seen, four Rabi flops can be made in the second microwave zone, without significant decrease of coherence. The required power to make a $\pi/2$ pulse in the first microwave zone is three times larger than that in the second microwave zone. This is attributed to a poor contact for the microwave incoupling. For the Ramsey-type measurements presented in the next sections, we balance the power in the two microwave zones by adding an attenuator to the cable that feeds the second microwave zone.

\begin{figure}
\begin{center}
\includegraphics[width=\linewidth]{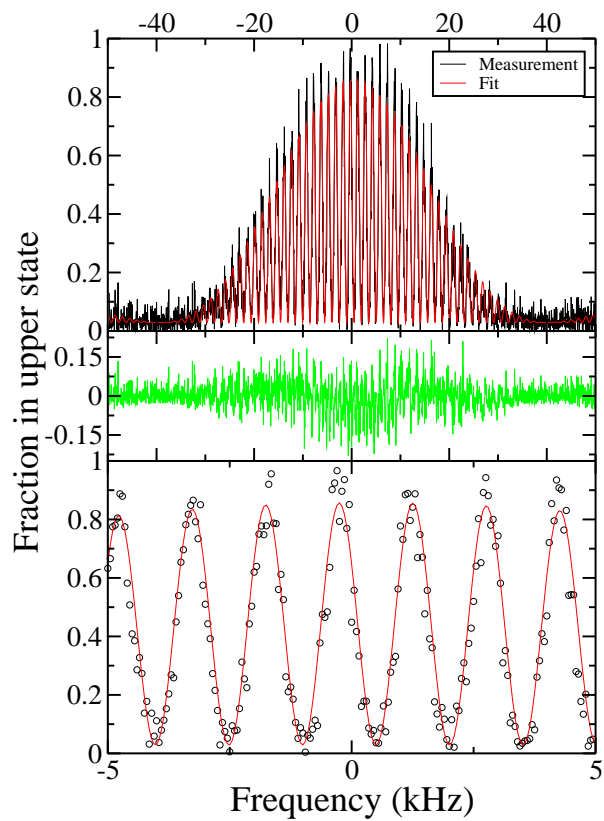}
\end{center}
\caption{Ramsey-type measurement of the  $m_J^B=+1 \rightarrow m_J^B=+1$ transition. The black curve in the top panel shows the measured ratio of the number of molecules in the initial and final states, while the red curve results from a fit to the data. The green curve in the middle panel shows the difference between the experimental data and the fit. The lower panel shows a zoomed in part around the resonance frequency. The frequency axis is offset by 394~229~829~Hz.}
\label{fig:specramsey}
\end{figure}

The linewidth of the resonance recorded by a single microwave zone is limited by the interaction time, $\Delta f \approx 1/\tau = v/l$, where $v$ is the velocity of the molecular beam and $l$ is the length of the microwave zone. In our case this corresponds to about 40~kHz. In order to decrease the linewidth and thereby increase the accuracy of the experiment, one needs to use slower molecules or a longer microwave zone. Ramsey demonstrated a more elegant way to reduce the linewidth by using two separate microwave zones~\cite{Ramsey:Nobel}. In the first microwave zone a $\pi/2$ pulse is used to create an equal superposition of the upper and lower level. While the molecules are in free flight from the first to the second zone, the phase between the two coefficients that describe the superposition evolves at the transition frequency. In the second microwave zone, this phase is probed using another $\pi/2$ pulse. If the frequency of the microwave field that is applied to the microwave zones is equal to the transition frequency, the second pulse will be in phase with the phase evolution of the superposition and all molecules will end up in the excited state. If the frequency is however slightly different, the second pulse will be out of phase with the phase evolution of the superposition, and only a fraction of the molecules end up in the excited state. The so-called Ramsey fringes that appear when the frequency of the microwave field is scanned, have a periodicity that is now given by $v/L$, where $L$ is the distance between the two microwave zones.       

The black curve in the top panel of Fig.~\ref{fig:specramsey} shows the result of a Ramsey-type measurement of the transition from $m_J^B=+1$ in the lower lambda-doublet component to $m_J^B=+1$ in the upper lambda-doublet component. Ramsey-fringes appear as a rapid cosine modulation on a broad sinc line shape. The width of the sinc is determined by the interaction time from each microwave zone separately, and is $\sim v/l$=800/0.02=40~kHz. The period of the cosine modulation is determined by the flight time between the two zones and is $v/L$=800/0.5=1.6~kHz. The red curve results from a fit to the data using 

\begin{equation}
y=a_{0}+a_{1}\mathrm{sinc}^{2}(a_2 x)\cos^2(\pi x/a_3).
\label{eq:Ramsey}
\end{equation} 

\noindent
The green curve in the middle panel shows the difference between the experimental data and the fit. The bottom panel shows a zoomed in part of the central part of the peak. The black data points show the experimental data while the red curve is a fit. Note that, strictly speaking, Eq.~\ref{eq:Ramsey} is only valid in the case of weak excitation; a more correct lineshape is given in~\cite{Ramsey:Book}. The observed deviations between the data and the fit are attributed to the fact that a fraction of the molecules that are deflected hit the lower electrode and are lost from the beam. A better fit is obtained by adding a sin$^4$ term. As the deviations are symmetric around zero, the obtained transition frequency is not affected.  

\begin{figure}
\begin{center}
\includegraphics[width=\linewidth]{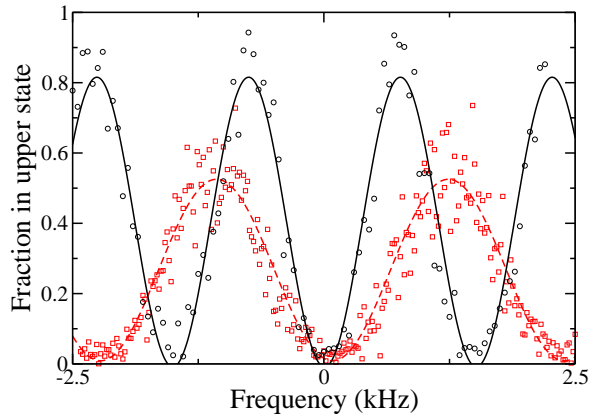}
\end{center}
\caption{A recording of the central fringe in a beam of pure CO (black circles) and a beam of CO seeded in helium (red squares). A difference between the frequencies of the central fringes can be observed. This is due to phase shifts, and it can be eliminated by extrapolating to zero velocity.}
\label{fig:centralfringe}
\end{figure}

In order to identify the central fringe, we have recorded the fringe pattern using beams at different velocities. Fig.~\ref{fig:centralfringe} shows the central fringe recorded in a beam of pure CO (black circles) and a beam of CO seeded in helium (red squares). The curves also shown are a sin$^2$ fit to the data points. As a result of the higher velocity of the CO in helium, the observed fringes are wider. The central fringe, however, is always found near (but not exactly at) the transition frequency. Note that in our setup, the two inner field plates are grounded. As a result the two zones have a $\pi$ phase difference and the transition frequency corresponds to a minimum in the fringe pattern. As observed, there is a small frequency shift between the measurements due to a phase difference between the microwave zones. The true transition frequency is found by extrapolating to zero velocity.

To determine the transition frequency, we typically record two fringes around the central fringe and fit a sin$^2$ function to the data. Such a scan takes approximately 600~seconds and allows us to determine the central frequency with an statistical uncertainty of about 4~Hz for measurements in a pure beam of CO and about 8~Hz for measurements on CO seeded in helium. These uncertainties are close to the ones expected from the number of molecules that are detected in our measurements. Note that by simultaneously measuring the number of molecules in the initial and the final state, shot-to-shot noise from the pulsed beam and the laser is canceled. Hence, we expected to be limited by quantum projection noise only. Indeed, on time scales below a few minutes the uncertainty reaches the shot-noise limit. On longer time scales, however, the statistical uncertainty is larger than expected. We attribute this to fluctuations of the magnetic bias field (\textit{vide infra}). 

In order to quantify possible systematic effects, we have recorded many single scans while varying all parameters that may influence the transition frequency.    

\begin{figure}
\begin{center}
\includegraphics[width=\linewidth]{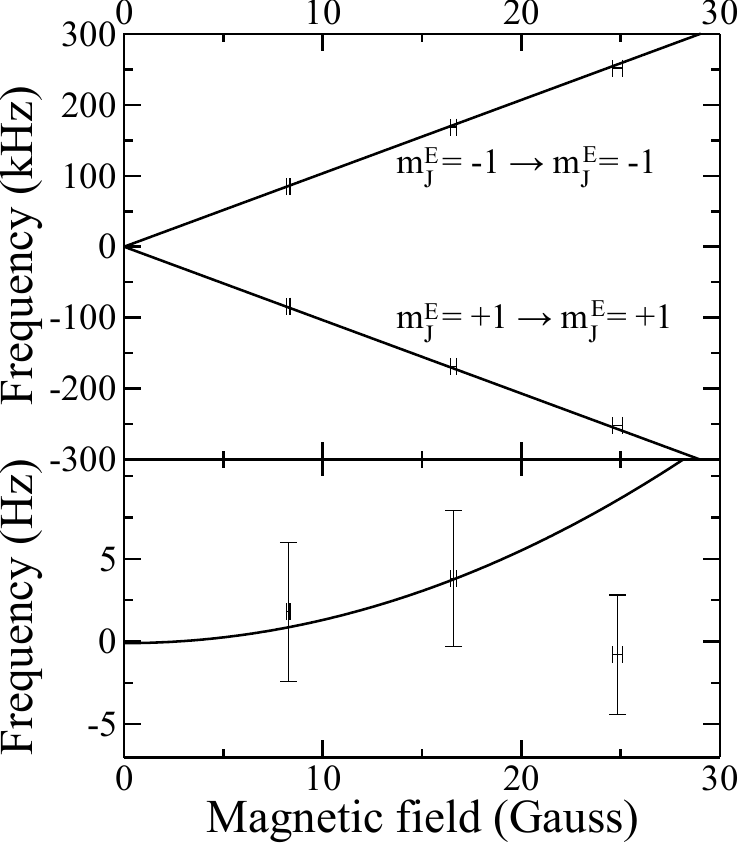}
\end{center}
\caption{Recorded frequencies of the $m_J^B = -1 \rightarrow m_J^B = -1$ and the $m_J^B = +1 \rightarrow m_J^B = +1$ transitions as a function of the magnetic field (upper panel) and the averaged value of these two transitions (lower panel). The solid curve results from a calculation using \textsc{Pgopher}. Both frequency axes are offset by 394~064~983.6~Hz.}
\label{fig:zeemanshift}
\end{figure}

\textit{Linear and quadratic Zeeman shift}. The frequency that we obtain from a single measurement depends on the strength of the magnetic bias field as the recorded transitions experience a (differential) linear Zeeman shift. The upper panel of Fig.~\ref{fig:zeemanshift} shows the $m_J^B=+1 \rightarrow m_J^B=+1$ and $m_J^B=-1 \rightarrow m_J^B=-1$ transitions at three different magnetic fields. The $x$-axis displays the average magnetic field over the flight path of the molecules. As observed, the $m_J^B=+1 \rightarrow m_J^B=+1$ and $m_J^B=-1 \rightarrow m_J^B=-1$ transitions experience an equal but opposite linear Zeeman shift of $\sim$10~kHz/Gauss. The solid line results from a calculation with \textsc{pgopher}~\cite{pgopher} using the molecular constants from de Nijs \textit{et al.}~\cite{denijs:pra2011} and the g-factor from Havenith \textit{et al.}~\cite{Havenith:MolPhys1988}. In order to determine the field free transition frequency, we take the average of the $m_J^B=+1 \rightarrow m_J^B=+1$ and $m_J^B=-1 \rightarrow m_J^B=-1$ transitions recorded at the same magnetic field. These averages are shown in the lower panel of Fig.~\ref{fig:zeemanshift}. The solid line shows a calculation with \textsc{pgopher}. From this calculation, we find that the linear Zeeman effect cancels exactly while the quadratic Zeeman shift is 14~mHz/Gauss$^2$. We have performed most of our measurements at a bias magnetic field of $\sim$17 Gauss; at this field the quadratic Zeeman shift is still only 4~Hz while depolarization of the beam is avoided.
  
\textit{Magnetic field instabilities}. Due to the sensitivity of the two transitions to the strength of the magnetic field, any magnetic field noise is translated into frequency noise. As a result of these fluctuations, the decrease of the uncertainty in our measurement as a function of measurement time is smaller than expected from the number of molecules that are detected. Whereas fluctuations of the magnetic field on short timescales add noise,  fluctuations on longer time scales may give rise to systematic shifts. In order to cancel slow drifts, we switch between the $m_J^B=+1 \rightarrow m_J^B=+1$ and $m_J^B=-1 \rightarrow m_J^B=-1$ transitions every 20 minutes, limited by the time it takes to change the frequency and polarization of the UV laser.  

\begin{figure}
\begin{center}
\includegraphics[width=\linewidth]{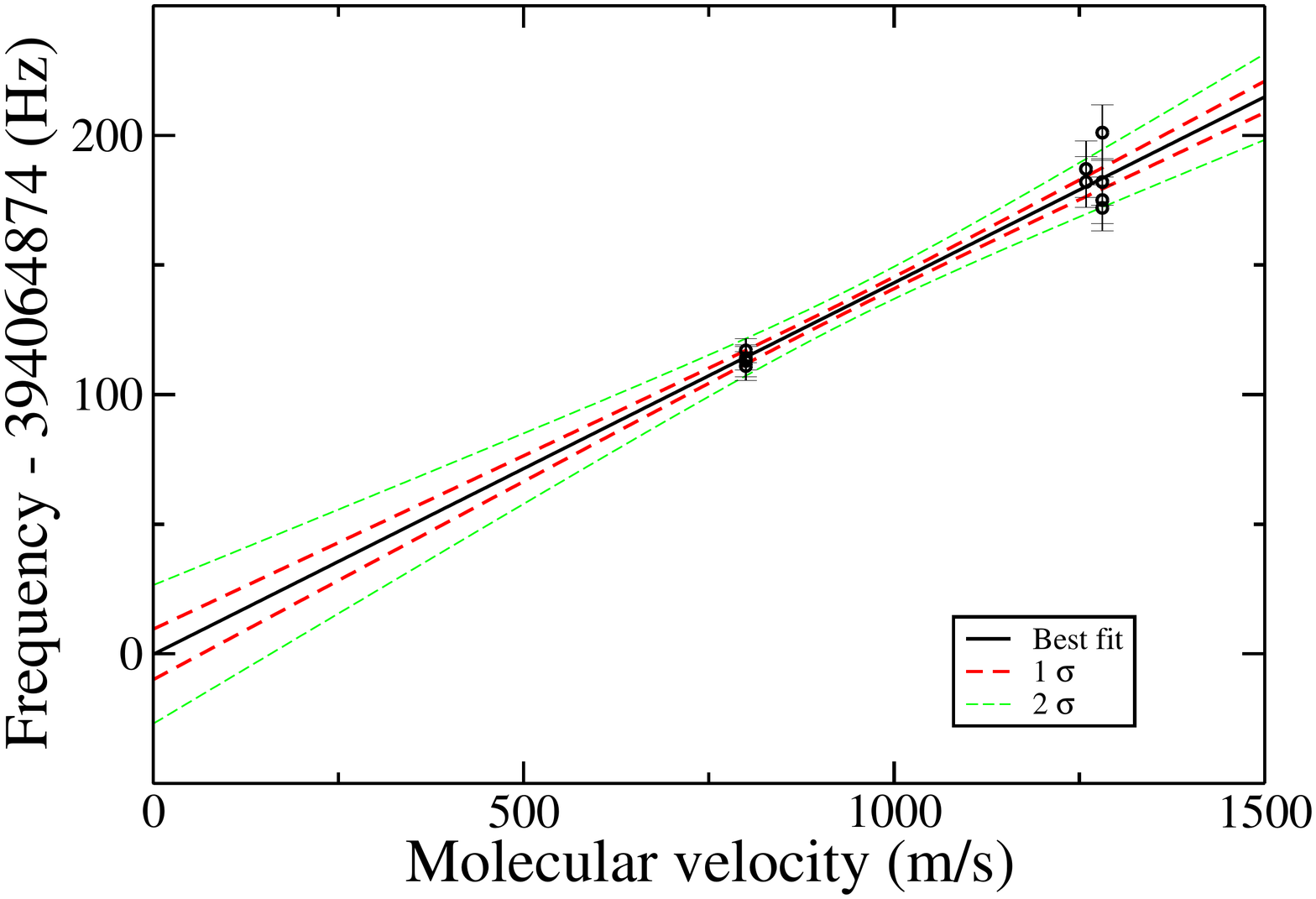}
\end{center}
\caption{Measurements of the transition frequency at different molecular beam velocities. The black, solid, curve shows a fitted linear slope, the red and green dashed curves show one and two standard deviation confidence bands, respectively.}
\label{fig:extrapolation}
\end{figure}

\textit{Phase offsets}. If the cables that connect the two microwave zones are not of the same length, there will be a phase shift that gives rise to a velocity dependent frequency shift. As in our setup a difference in length as small as 1~mm results in a frequency shift of 3~Hz at 800~m/s, the phase shift needs to be measured directly. Therefore we have recorded the transitions at different molecular velocities and extrapolate to zero, as shown in Fig.~\ref{fig:extrapolation}. From a total of 20 scans, of the $m_J^B=+1 \rightarrow m_J^B=+1$ and $m_J^B=-1 \rightarrow m_J^B=-1$ transitions measured in a beam of pure CO, $v$=800~m/s, and in a beam of 20\% CO in helium, $v$=1270~m/s, we find the extrapolated frequency to be equal to 394~064~874(10) Hz. The velocity of the molecular beam is determined from the known dimensions of the molecular beam machine and the time-delay between the pulsed excitation laser and the gate-pulse applied to the detector. Note that uncertainties in the distance between the excitation zone and the detector are canceled when we extrapolate to zero velocity. In order to find the field-free transition, we have to account for a 4~Hz shift due to the the quadratic Zeeman shift in the magnetic bias field. Thus, finally the true transition frequency is found to be 394~064~870(10) Hz.

\textit{2nd order Doppler shift and motional Stark effect}. The shift due to the motional Stark effect is estimated to be
below 1~Hz. Moreover, it scales linearly with the velocity and is compensated by extrapolating to zero velocity. The second order Doppler is $\Delta \nu = \nu_0 (v/c)^2 \approx 3$~mHz and is negligible at the accuracy level of the experiment.

\begin{figure}
\begin{center}
\includegraphics[width=\linewidth]{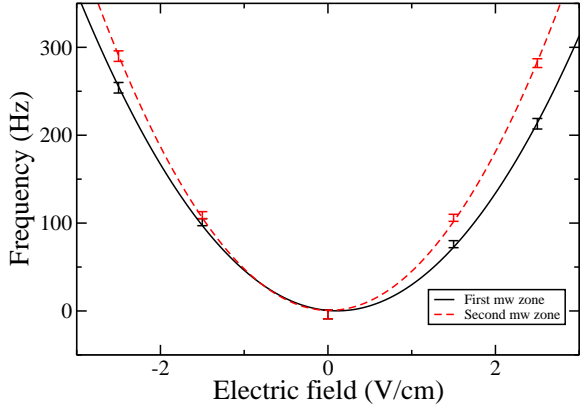}
\end{center}
\caption{Recorded frequencies of the $m_J^B = -1 \rightarrow m_J^B = -1$ transition as a function of the applied DC voltage in the first (black circles) and second (red squares) microwave zone. The solid black and dashed red curves result from quadratic fits to the data.}
\label{fig:starkshift}
\end{figure}

\textit{DC-Stark shift}. Electric fields in the interaction region due to leakage from the excitation and deflection fields, patch potentials in the tube and contact potentials between the field plates of the microwave zone induce Stark shifts. As the Stark shifts in the $m_J^B = -1 \rightarrow m_J^B = -1$ and the $m_J^B = +1 \rightarrow m_J^B = +1$ transitions are equal and both positive, they are not canceled by taking the average of the two. The biggest effect may be expected as a result of contact potentials. We have tested possible Stark shifts in the microwave zones by adding a small DC component to the microwave field. Fig.~\ref{fig:starkshift} shows the resonance frequency of the $m_J^B = -1 \rightarrow m_J^B = -1$ transition as a function of the applied DC voltage in the first (black data points) and second (red data points) microwave zones. The black and red curves show a quadratic fit to the data. The Stark shift is almost symmetric around zero, the residual DC-Stark shift is estimated to be below 1~Hz.

\textit{ac-Stark shift}. To study the effect of the microwave power on the transition frequency, we have 
varied the microwave power by over an order of magnitude. No significant dependence of the frequency on microwave power was found.

\textit{Absolute frequency determination}. The microwave source used in this work (Agilent E8257D) is linked to a Rubidium clock and has an absolute frequency uncertainty of 10$^{-12}$, i.e., $\sim$0.4~mHz at 394~MHz, negligible at the accuracy level of the experiment.

\section{Conclusion}

Using a Ramsey-type setup, the lambda-doublet transition in the $J=1,\, \Omega=1$ level of the \api\ state of CO was measured to be 394~064~870(10) Hz. Frequency shifts due to phase shifts between the two microwave zones of the Ramsey spectrometer are canceled by recording the transition frequency at different velocities and extrapolating to zero velocity. Our measurements are performed in a magnetic bias field of $\sim$17~Gauss. Frequency shifts due to the linear Zeeman effect in this field are canceled by taking the average of the $m_J^B = +1 \rightarrow m_J^B = +1$ and $m_J^B = -1 \rightarrow m_J^B = -1$ transitions. The quadratic Zeeman effect gives rise to a shift of 4~Hz which is taken into account in the quoted transition frequency. Other possible systematic frequency shifts may be neglected within the accuracy of the measurement. The obtained result is in agreement with measurements by Wicke~\textit{et al.}~\cite{Wicke:1972}, but is a 100 times more accurate.

An important motivation for this work is to estimate the possible accuracy that might be obtained on the two-photon transition connecting the $J=6,\, \Omega=1$ level to the $J=8,\, \Omega=0$ level that is exceptionally sensitive to a possible time-variation of the fundamental constants~\cite{Bethlem:FarDisc,denijs:pra2011}. An advantage of this transition is that the $m_J^B = 0 \rightarrow m_J^B = 0$ transition can be measured directly, thus avoiding the problems with the stability of the magnetic bias field. A disadvantage is that the population in the $J=5$ level is much smaller than the population in the $J=1$ level, reducing the number of molecules that is observed, while the Stark shift is considerable less, making it necessary to use a longer deflection field. 

In order to obtain a constraint on the time-variation of $\mu$ at the level of 5.6 $\times$ 10$^{-14}$/yr, the current best limit set by spectroscopy on SF$_6$~\cite{shelkovnikov:prl}, we would need to record the $J=6,\, \Omega=1 \rightarrow J=8,\, \Omega=0$ two-photon transition at 1.6\,GHz  with an accuracy 0.03~Hz over an interval of one year. To reach this precision within a realistic measurement time, say 24 hours, the number of detected molecules should be at least 2500 per shot, i.e., 50 times more than detected in the current experiment but now starting from the less populated $J=5$ level. Although challenging, this seems possible by using a more efficient detector~\cite{Bethlem:prl1999} and/or by using quadrupole or hexapole lenses to collimate the molecular beam~\cite{Jongma:JCP}.

\section{Acknowledgements}
We thank Laura Dreissen (VU Amsterdam) for help with the experiments and Leo Meerts (RU Nijmegen) and Stefan Truppe (Imperial College London) for helpful discussions. This work is financially supported by the Netherlands Foundation for Fundamental Research of Matter (FOM) (project 10PR2793 and program ``Broken mirrors and drifting constants''). W.U. acknowledges support from the Templeton Foundation. H.L.B. acknowledges support from NWO via a VIDI-grant.


\end{document}